\def\year{2015}
\documentclass[letterpaper]{article}
\usepackage{aaai}
\usepackage{times}
\usepackage{helvet}
\usepackage{courier}
\usepackage{amsmath}
\usepackage{amssymb}
\usepackage{paralist}
\usepackage{graphicx}
\usepackage{subcaption}

\newcommand{\vocab}{\emph}

\renewcommand\_{\textunderscore\allowbreak}

\begin{document}
%
\title{Eliciting Disease Data from Wikipedia Articles}
\author{Geoffrey Fairchild\\
\Large{\textbf{Sara Y. Del Valle}}\\
Los Alamos National Laboratory\\
Defense Systems \& Analysis Division\\
Los Alamos, New Mexico, USA\\
\And
Lalindra De Silva\\
The University of Utah\\
School of Computing\\
Salt Lake City, Utah, USA\\
\And
Alberto M. Segre\\
The University of Iowa\\
Department of Computer Science\\
Iowa City, Iowa, USA}
\maketitle
\begin{abstract}
\begin{quote}
Traditional disease surveillance systems suffer from several disadvantages, including reporting lags and antiquated technology, that have caused a movement towards internet-based disease surveillance systems. Internet systems are particularly attractive for disease outbreaks because they can provide data in near real-time and can be verified by individuals around the globe. However, most existing systems have focused on disease monitoring and do not provide a data repository for policy makers or researchers. In order to fill this gap, we analyzed Wikipedia article content. 

We demonstrate how a named-entity recognizer can be trained to tag case counts, death counts, and hospitalization counts in the article narrative that achieves an F1 score of 0.753. We also show, using the 2014 West African Ebola virus disease epidemic article as a case study, that there are detailed time series data that are consistently updated that closely align with ground truth data.

We argue that Wikipedia can be used to create the first community-driven open-source emerging disease detection, monitoring, and repository system.
\end{quote}
\end{abstract}

\section{Introduction}

Most traditional disease surveillance systems rely on data from patient visits or lab records~\cite{Losos1996,Burkhead2000,Adams2013}. These systems, while generally recognized to contain accurate information, rely on a hierarchy of public health systems that causes reporting lags of up to 1--2 weeks in many cases~\cite{Burkhead2000}. Additionally, many regions of the world lack the infrastructure necessary for these systems to produce reliable and trustworthy data. Recently, in an effort to overcome these issues, timely global approaches to disease surveillance have been devised using internet-based data. Data sources such as search engine queries (e.g., \cite{Polgreen2008,Ginsberg2009}), Twitter (e.g., \cite{Culotta2010,Aramaki2011,Paul2011,Signorini2011}), and Wikipedia access logs (e.g., \cite{McIver2014,Generous2014}) have been shown to be effective in this arena.

A notably different internet-based disease surveillance tool is HealthMap~\cite{Freifeld2008}. HealthMap analyzes, in real-time, data from a variety of sources (e.g., ProMED-mail~\cite{Madoff2004}, Google News, the World Health Organization) in order to allow simple querying, filtering, and visualization of outbreaks past and present. During emerging outbreaks, HealthMap is often used to understand the current state (e.g., incidence and death counts, outbreak locations). For example, HealthMap was able to detect the 2014 Ebola epidemic nine days before the World Health Organization (WHO) officially announced it~\cite{Greenemeier2014}.

While HealthMap has certainly been influential in the digital disease detection sphere, it has some drawbacks. First and foremost, it runs on source code that is not open and relies on certain data sources that are not freely available in their entirety (e.g., Moreover Newsdesk\footnote{http://www.moreover.com/}). Some argue that there is a genuine need for open source code and open data in order to validate, replicate, and improve existing systems~\cite{Generous2014}. They argue that while certain closed source services, such as HealthMap and Google Flu Trends~\cite{Ginsberg2009}, are popular and useful to the public, there is no way for the public to contribute to the service or continue the service, should the owners decide to shut it down. For example, Google offers a companion site to Google Flu Trends, Google Dengue Trends\footnote{http://www.google.org/denguetrends/}. However, since Google's source code and data are closed, it is not possible for anyone outside of Google to create similar systems for other diseases, e.g., Google Ebola Trends. Additionally, it is not possible for anyone outside of the HealthMap development team to add new features or data sources to HealthMap. For these reasons, Generous et al. argue for the use of Wikipedia access logs coupled with open source code for digital disease surveillance.

Much richer Wikipedia data are available, however, than just access logs. The entire Wikipedia article content and edit histories are available, complete with edit history metadata (e.g., timestamps of edits and IP addresses of anonymous editors). A plethora of open media---audio, images, and video---are also available.

Wikipedia has a history of being edited and used, in many cases, in near real-time during unfolding news events. Keegan et al. have been particularly instrumental in understanding Wikipedia's dynamics during unfolding breaking news events, such as natural disasters and political conflicts and scandals~\cite{Keegan2011,Keegan2013,Keegan2013a}. They have provided insight into editor networks as well as editing activity during news events. Recognizing that Wikipedia might offer useful disease data during unfolding epidemiological events, this study presents a novel use of Wikipedia article content and edit history in which disease data (i.e., case, death, and hospitalization counts) are elicited in a timely fashion.

We study two different aspects of Wikipedia content as it relates to unfolding disease events:
\begin{enumerate}
	\item Using standard natural language processing (NLP) techniques, we demonstrate how to capture case counts, death counts, and hospitalization counts from the article text.
	\item Using the 2014 West African Ebola virus epidemic article as a case study, we show there are valuable time series data present in the tables found in certain articles.
\end{enumerate}
We argue that Wikipedia data can not only be used for disease surveillance but also as a centralized repository system for collecting disease-related data in near real-time.

\section{Methods}

Disease-related information can be found in a number of places on Wikipedia. We demonstrate how two aspects of Wikipedia article content (historical changes to article text and tabular content) can be harvested for disease surveillance purposes. We first show how a named-entity recognizer can be trained to elicit ``important'' phrases from outbreak articles, and we then study the accuracy of tabular time series data found in certain articles using the 2014 West African Ebola epidemic as a case study.

\subsection{Wikipedia data}

Wikipedia is an open collaborative encyclopedia consisting of approximately 30 million articles across 287 languages~\cite{Wikimedia2014_Wikipedia_Wikipedia,Wikimedia2014_Wikipedia_statistics}. The English edition of Wikipedia is by far the largest and most active edition; it alone contains approximately 4.7 million articles, while the next largest Wikipedia edition (Swedish) contains only 1.9 million articles~\cite{Wikimedia2014_Wikipedia_statistics}. The textual content of the current revision of each English Wikipedia article totals approximately 10 gigabytes~\cite{Wikimedia2014_English_Wikipedia}.

One of Wikipedia's primary attractions to researchers is its openness. All of the historical article content, dating back to Wikipedia's inception in 2001, is available to anyone free of charge. Wikipedia content can be acquired through two means:
\begin{inparaenum}[\itshape a\upshape)]
	\item Wikipedia's official web API\footnote{http://www.mediawiki.org/wiki/API:Main\_page} or
	\item downloadable database dumps\footnote{http://dumps.wikimedia.org/enwiki/latest/}.
\end{inparaenum}
Although the analysis in this study could have been done offline using the downloadable database dumps, this option is in practice difficult, as the database dumps containing all historical English article revisions are very large (multiple terabytes when uncompressed)~\cite{Wikimedia2014_Wikipedia_database_download}. We therefore decided to use Wikipedia's web API, caching content when appropriate.

Wikipedia contains many articles on specific disease outbreaks and epidemics (e.g., the 2014 West Africa Ebola epidemic\footnote{http://en.wikipedia.org/wiki/Ebola\_virus\_epidemic\_in\_West\_Africa} and the 2012 Middle Eastern Respiratory Syndrome Coronavirus (MERS-CoV) outbreak\footnote{http://en.wikipedia.org/wiki/2012\_Middle\_East\_respiratory\_syndrome\_coronavirus\_outbreak}). We identified two key aspects of Wikipedia disease outbreak articles that can aid disease surveillance efforts:
\begin{inparaenum}[\itshape a\upshape)]
	\item key phrases in the article text and
	\item tabular content.
\end{inparaenum}
Most outbreak articles we surveyed contained: dates, locations, case counts, death counts, case fatality rates, demographics, and hospitalization counts in the text. These data are, in general, swiftly updated as new data become available. Perhaps most importantly, sources are often provided so that external review can occur. The following two excerpts came from the articles on the 2012 MERS-CoV outbreak and 2014 Ebola epidemic, respectively:
\begin{quote}
	On 16 April 2014, Malaysia reported its first MERS-COV related death.$^{[34]}$ The person was a 54 year-old man who had traveled to Jeddah, Saudi Arabia, together with pilgrimage group composed of 18 people, from 15--28 March 2014. He became ill by 4 April, and sought remedy at a clinic in Johor on 7 April. He was hospitalized by 9 April and died on 13 April.$^{[35]}$~\cite{Wikimedia2014_Wikipedia_MERS_outbreak}
\end{quote}
\begin{quote}
	On 31 March, the U.S. Centers for Disease Control and Prevention (CDC) sent a five-person team to assist Guinea's Ministry of Health and the WHO to lead an international response to the Ebola outbreak. On that date, the WHO reported 112 suspected and confirmed cases including 70 deaths. Two cases were reported from Liberia of people who had recently traveled to Guinea, and suspected cases in Liberia and Sierra Leone were being investigated.$^{[24]}$ On 30 April, Guinea's Ministry of Health reported 221 suspected and confirmed cases including 146 deaths. The cases included 25 health care workers with 16 deaths. By late May, the outbreak had spread to Conakry, Guinea's capital, a city of about two million inhabitants.$^{[24]}$ On 28 May, the total cases reported had reached 281 with 186 deaths.$^{[24]}$~\cite{Wikimedia2014_Wikipedia_Ebola_epidemic}
\end{quote}

\begin{figure}
	\centering
	\includegraphics[width=\columnwidth]{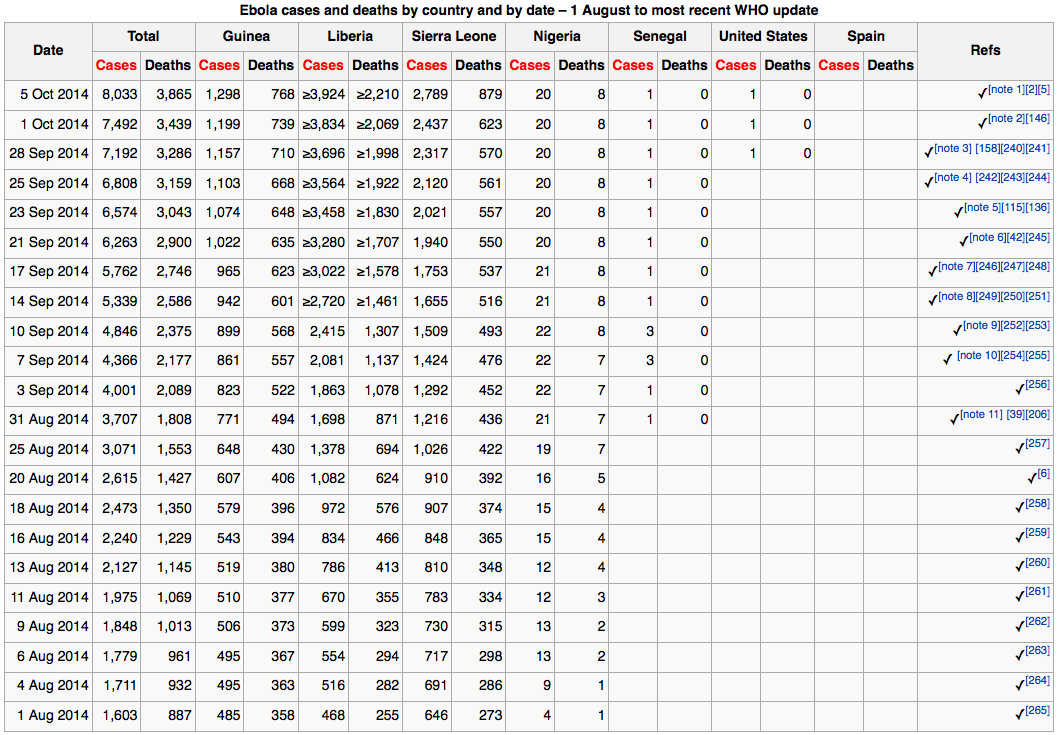}
	\caption{Table containing updated worldwide Ebola case counts and death counts. This is a screenshot taken directly from the 2014 Ebola epidemic Wikipedia article~\protect\cite{Wikimedia2014_Wikipedia_Ebola_epidemic}. Time granularity is irregular but is in general every 2--5 days. References are also provided for all data points.}
	\label{figure:ebola_table}
\end{figure}

Although most outbreak articles contain content similar to the above examples, not all outbreak articles on Wikipedia contain tabular data. The tabular data that do exist, though, are often consistently updated. For example, Figure~\ref{figure:ebola_table} presents a screenshot of a table taken from the 2014 Ebola epidemic article. This table contains case counts and death counts for all regions of the world affected by the epidemic, complete with references for the source data. The time granularity is irregular, but updated counts are consistently provided every 2--5 days.

While there are certainly other aspects of Wikipedia article content that can be leveraged for disease surveillance purposes, these are the two we focus on in this study. The following sections detail the data extraction methods we use.

\subsection{Named-entity recognition}

In order to recognize certain key phrases in the Wikipedia article narrative, we trained a \vocab{named-entity recognizer} (NER). Named-entity recognition is a task commonly used in natural language processing (NLP) to identify and categorize certain key phrases in text (e.g., names, locations, dates, organizations). NERs are \vocab{sequence labelers}; that is, they label sequences of words. Consider the following example~\cite{Wikimedia2014_Wikipedia_NER}:
\begin{quote}
	Jim bought 300 shares of Acme Corp. in 2006.
\end{quote}
Entities in this example could be named as follows:
\begin{quote}
	[Jim]$_{\mathtt{PERSON}}$ bought 300 shares of [Acme Corp.]$_{\mathtt{ORGANIZATION}}$ in [2006]$_{\mathtt{TIME}}$. 
\end{quote}

This study specifically uses Stanford's NER~\cite{Finkel2005}\footnote{http://nlp.stanford.edu/software/CRF-NER.shtml}. The Stanford NER is an implementation of a conditional random field (CRF) model~\cite{Sutton2011}. CRFs are probabilistic statistical models that are the discriminative analog of hidden Markov models (HMMs). Generative models, such as HMMs, learn the joint probability $p(x, y)$, while discriminative models, such as CRFs, learn the conditional probability $p(y \mid x)$. In practice, this means that generative models like HMMs classify by modeling the actual distribution of each class, while discriminative models like CRFs classify by modeling the boundaries between classes. In most cases, discriminative models outperform generative models~\cite{Ng2002}.

While Stanford's NER includes models capable of recognizing common named entities, such as \texttt{PERSON}, \texttt{ORGANIZATION}, and \texttt{LOCATION}, it also provides the capability for us to train our own model so that we can capture new types of named entities we are interested in. For this specific task, we were interested in automatically identifying three entity types:
\begin{inparaenum}[\itshape a\upshape)]
	\item \texttt{DEATHS}
	\item \texttt{INFECTIONS}, and
	\item \texttt{HOSPITALIZATIONS}.
\end{inparaenum}
Our trained model should therefore be able to automatically tag phrases that correspond to these three entities in the text documents it receives as input.

NERs possess the ability to learn and generalize in order to identify unseen phrase patterns. Since the classifier is dependent on the features we provide to it (e.g., words, part of speech tags), it should hopefully generalize well for the unseen instances. A more simplistic pattern-matching approach, such as regular expressions, is not practical due to inherent variation. For example, the following phrases from our dataset all contain \texttt{INFECTIONS} entities:
\begin{enumerate}
	\item \ldots a total of 17 patients with confirmed H7N9 virus infection \ldots
	\item \ldots there were only sixty-five cases and four deaths \dots
	\item \ldots more than 16,000 cases were being treated \dots
\end{enumerate}
Example 1 has the pattern \texttt{[number] patients}, while examples 2 and 3 follow the pattern \texttt{[number] cases}. However, example 2 spells out the number, while example 3 provides the numeral. A simple regular expression cannot capture the variability found in our dataset; we would need to define dozens of regular expressions for each entity type, and rigidity of regular expressions would limit the likelihood that we would be able to identify entities in new unseen patterns.

A number of steps were required to prepare the data for annotation so that the NER could be trained:
\begin{enumerate}
	\item We first queried Wikipedia's API in order to get the complete revision history for the articles used in our training set.
	\item We cleaned each revision by stripping all MediaWiki markup from the text, as well as removing tables.
	\item We computed the diff (i.e., textual changes) between successive pairs of articles. This provided lines deleted and added between the two article revisions. We retained a list of all the line additions across all article revisions.
	\item Many lines in this resulting list were similar to one another (e.g., ``There are 45 new cases." $\rightarrow$ ``There are 56 new cases."). For the purposes of training the NER, it is not necessary to retain highly similar or identical lines. We therefore split each line into sentences and removed similar sentences by computing the Jaccard similarity between each sentence using trigrams as the constituent parts in the Jaccard equation. The Jaccard similarity equation for measuring the similarity between two sets $A$ and $B$, defined as $J(A,B) = \frac{\lvert A \cap B \rvert}{\lvert A \cup B \rvert}$, is commonly used for near-duplicate detection~\cite{Manning2009}. We only kept sentences for which the similarity with all the distinct sentences retained so far was no greater than 0.75.
	\item We split each line into tokens in order to create a tab-separated value file that is compatible with Stanford's NER.
	\item Finally, we used Stanford's part-of-speech (POS) tagger~\cite{Toutanova2003}\footnote{http://nlp.stanford.edu/software/tagger.shtml} to add a POS feature to each token.
\end{enumerate}

In order to train the NER, we annotated a dataset derived from the following 14 Wikipedia articles generated according to the above methodology:
\begin{inparaenum}[\itshape a\upshape)]
	\item Ebola virus epidemic in West Africa\footnote{http://en.wikipedia.org/wiki/Ebola\_virus\_epidemic\_in\_West\_Africa},
	\item Haiti cholera outbreak\footnote{http://en.wikipedia.org/wiki/Haiti\_cholera\_outbreak},
	\item 2012 Middle East respiratory syndrome coronavirus outbreak\footnote{http://en.wikipedia.org/wiki/2012\_Middle\_East\_respiratory\_syndrome\_coronavirus\_outbreak},
	\item New England Compounding Center meningitis outbreak\footnote{http://en.wikipedia.org/wiki/New\_England\_Compounding\_Center\_meningitis\_outbreak},
	\item Influenza A virus subtype H7N9\footnote{http://en.wikipedia.org/wiki/Influenza\_A\_virus\_subtype\_H7N9},
	\item 2013--14 chikungunya outbreak\footnote{http://en.wikipedia.org/wiki/2013\%E2\%80\%9314\_chikungunya\_outbreak},
	\item Chikungunya outbreaks\footnote{http://en.wikipedia.org/wiki/Chikungunya\_outbreaks},
	\item Dengue fever outbreaks\footnote{http://en.wikipedia.org/wiki/Dengue\_fever\_outbreaks},
	\item 2013 dengue outbreak in Singapore\footnote{http://en.wikipedia.org/wiki/2013\_dengue\_outbreak\_in\_Singapore},
	\item 2011 dengue outbreak in Pakistan\footnote{http://en.wikipedia.org/wiki/2011\_dengue\_outbreak\_in\_Pakistan},
	\item 2009--10 West African meningitis outbreak\footnote{http://en.wikipedia.org/wiki/2009\%E2\%80\%9310\_West\_African\_meningitis\_outbreak},
	\item Mumps outbreaks in the 21st century\footnote{http://en.wikipedia.org/wiki/Mumps\_outbreaks\_in\_the\_21st\_century},
	\item Zimbabwean cholera outbreak\footnote{http://en.wikipedia.org/wiki/Zimbabwean\_cholera\_outbreak}, and
	\item 2006 dengue outbreak in India\footnote{http://en.wikipedia.org/wiki/2006\_dengue\_outbreak\_in\_India}.
\end{inparaenum}
The entire cleaned and annotated dataset contained approximately 55,000 tokens. The inside-outside-beginning (IOB) scheme, popularized in part by the CoNLL-2003 shared task on language-independent named-entity recognition~\cite{TjongKimSang2003}, was used to tag each token. The IOB scheme offers the ability to tie together sequences of tokens that make up an entity.

The annotation task was split between two annotators (the first and second authors). In order to tune inter-annotator agreement, the annotators each annotated three sets of 5,000 tokens. After each set of annotations, differences were identified, and clarifications to the annotation rules were made. The third set resulted in a Cohen's kappa coefficient of 0.937, indicating high agreement between the annotators.

\subsection{Tabular data}

To understand the viability of tabular data in Wikipedia, we concentrate on the Ebola virus epidemic in West Africa article\footnote{http://en.wikipedia.org/wiki/Ebola\_virus\_epidemic\_in\_West\_Africa}. We chose this article for two reasons. First, the epidemic is still unfolding, which makes it a concern for epidemiologists worldwide. Second, the epidemiological community has consistently updated the article as new developments are publicized. Ideally, we would analyze \emph{all} disease articles that contain tabular data, but the technical challenges surrounding parsing the constantly changing data leave this as future work.

Ebola is a rare but deadly virus that first appeared in 1976 simultaneously in two different remote villages in Africa. Outbreaks of Ebola virus disease (EVD), previously known as Ebola hemorrhagic fever (EHF), are sporadic and generally short-lived. The average case fatality rate is 50\%, but it has varied between 25\% and 90\% in previous outbreaks. EVD is transmitted to humans from animals (most commonly, bats, apes, and monkeys) and also from other humans through direct contact with blood and body fluids. Signs and symptoms appear within 2--21 days of exposure (average 8--10 days) and include fever, severe headache, muscle pain, weakness, diarrhea, vomiting, abdominal pain, and unexplained bleeding or bruising. Although there is currently no known cure, treatment in the form of aggressive rehydration seems to improve survival rates~\cite{WHO2014_Ebola,CDC2014_Ebola}.

The West African EVD epidemic was officially announced by the WHO on March 25, 2014~\cite{WHO2014_Ebola_announcement}. The disease spread rapidly and has proven difficult to contain in several regions of Africa. At the time of this writing, it has spread to 7 different countries (including two outside of Africa): Guinea, Liberia, Sierra Leone, Nigeria, Senegal, United States, and Spain.

\begin{figure}
	\centering
	\includegraphics[width=\columnwidth]{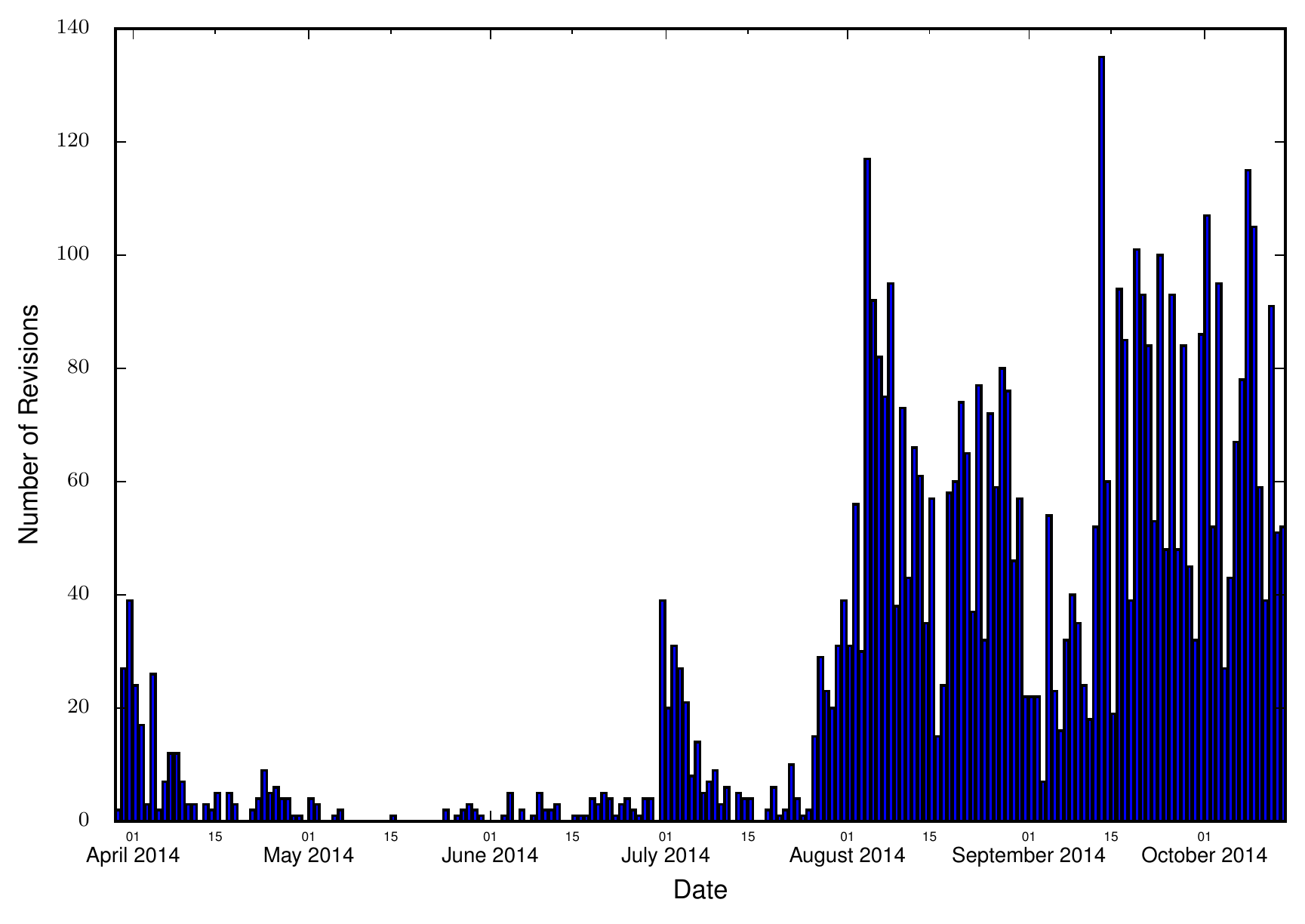}
	\caption{The number of revisions made each day to the 2014 Ebola virus epidemic in West Africa Wikipedia article (http://en.wikipedia.org/wiki/Ebola\_virus\_epidemic\_in\_West\_Africa). A total of 5,137 revisions were made over the 165-day period analyzed.}
	\label{figure:ebola_article_revisions}
\end{figure}

The Wikipedia article was created on March 29, 2014, four days after the WHO announced the epidemic~\cite{Wikimedia2014_Wikipedia_Ebola_epidemic_original}. As seen in Figure~\ref{figure:ebola_table}, this article contains detailed tables of case counts and death counts by country. The article is regularly updated by the Wikipedia community (see Figure~\ref{figure:ebola_article_revisions}); over the 165-day period analyzed, the article averaged approximately 31 revisions per day.

We parsed the Ebola article's tables in several steps:
\begin{enumerate}
	\item We first queried Wikipedia's API to get the complete revision history for the West African EVD epidemic article. Our initial dataset contained 5,137 revisions from March 29, 2014 to October 14, 2014.
	\item We then parsed each revision to pull out case count and death count time series for each revision. To parse the tables, we first used pandoc\footnote{http://johnmacfarlane.net/pandoc/} to convert the MediaWiki markup to consistently formatted HTML and then used BeautifulSoup\footnote{http://www.crummy.com/software/BeautifulSoup/} to parse the HTML. Because the Wikipedia time series contain a number of missing data points prior to June 30, 2014, we use this date for the beginning of our analysis; time series data prior to June 30, 2014 are not used in this study. This resulting dataset contained 3,803 time series.
	\item As Figure~\ref{figure:ebola_table} shows, there are non-regular gaps in the Wikipedia time series; these gaps range from 2--5 days. We used linear interpolation to fill in missing data points where necessary so that we have daily time series. Daily time series data simplify comparisons with ground truth data (described later).
	\item Recognizing that the tables will not necessarily change between article revisions (i.e., an article revision might contain edits to only the text of the article, not to a table in the article), we then removed identical time series. This final dataset contained 39 time series.
\end{enumerate}

\section{Results}

\subsection{Named-entity recognition}

To test the classifier's performance, we averaged precision, recall, and F1 score results from 10-fold cross-validation. Table~\ref{table:confusion_matrix} demonstrates a typical confusion matrix used to bin cross-validation results, which are then used to compute precision, recall, and the F1 score. Precision asks, ``Out of all the examples the classifier labeled, what fraction were correct?'' and is computed as $\frac{\mathrm{TP}}{\mathrm{TP} + \mathrm{FP}}$. Recall asks, ``Out of all labeled examples, what fraction did the classifier recognize?'' and is computed as $\frac{\mathrm{TP}}{\mathrm{TP} + \mathrm{FN}}$. The F1 score is the harmonic mean of precision and recall: $2 \cdot \frac{\mathrm{precision} \cdot \mathrm{recall}}{\mathrm{precision} + \mathrm{recall}}$. All three scores range from 0 to 1, where 0 is the worst score possible and 1 is the best score possible.

\begin{table}
	\centering
	\scriptsize
	\caption{Typical classifier confusion matrix.}
	\begin{tabular}{| l | l | l | l |}
		\hline
		& \textbf{Ground truth positive} & \textbf{Ground truth negative} \\ \hline
		\textbf{Test positive} & True positive (TP) & False positive (FP) \\ \hline
		\textbf{Test negative} & False negative (FN) & True negative (TN) \\ \hline
	\end{tabular}
	\label{table:confusion_matrix}
\end{table}

Table~\ref{table:cross-validation_results} shows these results as we varied the \texttt{maxNGramLeng} option (Stanford's default value is 6). The \texttt{maxNGramLeng} option determines sequence length when training. We were somewhat surprised to discover that larger \texttt{maxNGramLeng} values did not improve the performance of the classifier, indicating that more training data are likely necessary to further improve the classifier. Furthermore, roughly maximal performance is achieved with $\mathtt{maxNGramLeng}=4$; there is no tangible benefit to larger sequences (despite this, we concentrate on the $\mathtt{maxNGramLeng}=6$ case since it is the default). Our 14-article training set achieved precision of 0.812 and recall of 0.710, giving us an F1 score of 0.753 for $\mathtt{maxNGramLeng}=6$.

\begin{table}
	\centering
	\caption{Classifier performance determined from 10-fold cross-validation.}
	\begin{tabular}{| l | l | l | l |}
		\hline
		\textbf{\texttt{maxNGramLeng}} & \textbf{Precision} & \textbf{Recall} & \textbf{F1 score} \\ \hline
		1 & 0.820 & 0.693 & 0.747 \\ \hline
		2 & 0.810 & 0.690 & 0.740 \\ \hline
		3 & 0.815 & 0.702 & 0.750 \\ \hline
		4 & 0.814 & 0.709 & 0.753 \\ \hline
		5 & 0.813 & 0.709 & 0.753 \\ \hline
		6 & 0.812 & 0.710 & 0.753 \\ \hline
		7 & 0.812 & 0.706 & 0.751 \\ \hline
		8 & 0.814 & 0.708 & 0.753 \\ \hline
		9 & 0.815 & 0.707 & 0.753 \\ \hline
		10 & 0.815 & 0.708 & 0.753 \\ \hline
		11 & 0.813 & 0.708 & 0.753 \\ \hline
		12 & 0.811 & 0.709 & 0.752 \\ \hline
	\end{tabular}
	\label{table:cross-validation_results}
\end{table}

For $\mathtt{maxNGramLeng}=6$, Table~\ref{table:maxNGramLeng6_results} shows the average precision, recall, and F1 scores for each of the named entities we annotated (\texttt{DEATHS}, \texttt{INFECTIONS}, and \texttt{HOSPITALIZATIONS}). There were a total of 264 \texttt{DEATHS}, 633 \texttt{INFECTIONS}, and 16 \texttt{HOSPITALIZATIONS} entities annotated across the entire training dataset. Recall that we used the IOB scheme for annotating sequences; this is reflected in Table~\ref{table:maxNGramLeng6_results}, with \texttt{B-*} indicating the beginning of a sequence and \texttt{I-*} indicating the inside of a sequence. It is generally the case that identifying the beginning of a sequence is easier than identifying all of the inside words of a sequence; the only exception to this is \texttt{HOSPITALIZATIONS}, but we speculate that the identical beginning and inside results for this entity are due to the relatively small sample size.

\begin{table}
	\centering
	\small
	\caption{Classifier performance for each of the entities we used in our annotations.}
	\begin{tabular}{| l | l | l | l |}
		\hline
		\textbf{Named entity} & \textbf{Precision} & \textbf{Recall} & \textbf{F1 score} \\ \hline
		\texttt{B-Deaths} & 0.888 & 0.744 & 0.802 \\ \hline
		\texttt{I-Deaths} & 0.821 & 0.730 & 0.764 \\ \hline
		\texttt{B-Infections} & 0.812 & 0.719 & 0.756 \\ \hline
		\texttt{I-Infections} & 0.762 & 0.714 & 0.730 \\ \hline
		\texttt{B-Hospitalizations} & 0.933 & 0.833 & 0.853 \\ \hline
		\texttt{I-Hospitalizations} & 0.933 & 0.833 & 0.853 \\ \hline
	\end{tabular}
	\label{table:maxNGramLeng6_results}
\end{table}

\subsection{Tabular data}

To compute the accuracy of the Wikipedia West African EVD epidemic time series, we used Caitlin Rivers' crowdsourced Ebola data\footnote{https://github.com/cmrivers/ebola}. Her country-level data come from official WHO data and reports. As with the Wikipedia time series, we used linear interpolation to fill in missing data where necessary so that the ground truth data are specified daily; this ensured that the Wikipedia and ground truth time series were specified at the same granularity. Note that time granularity of the WHO-based ground truth dataset is generally finer than the Wikipedia data; the gaps in the ground truth time series were not the same as those in the Wikipedia time series. In many cases, the ground truth data were updated every 1--2 days.

\begin{figure}
	\centering
	\begin{subfigure}[b]{\columnwidth}
		\includegraphics[width=\columnwidth]{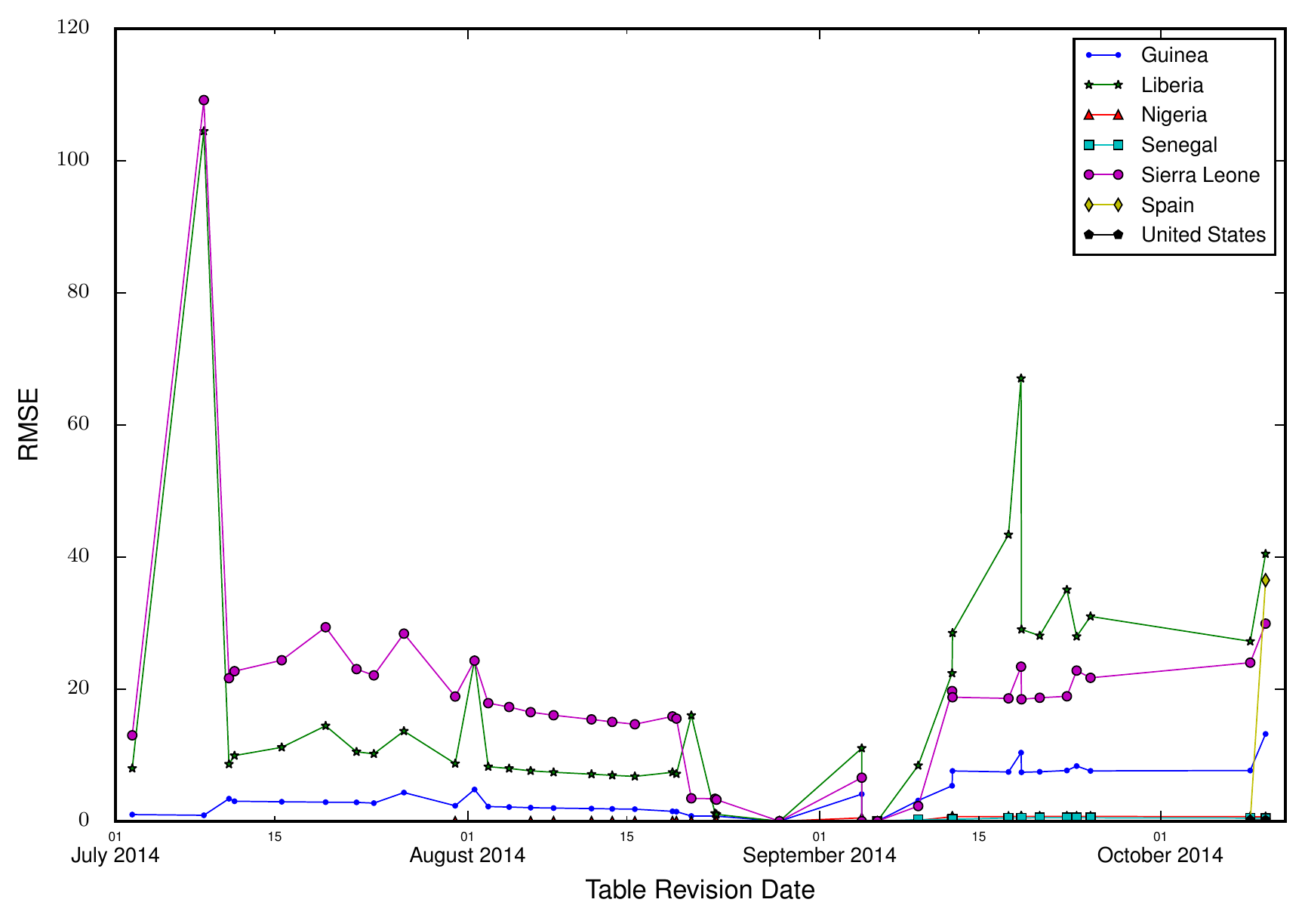}
		\caption{Cases}
        \label{figure:cases_rmse}
	\end{subfigure}
	\begin{subfigure}[b]{\columnwidth}
		\includegraphics[width=\columnwidth]{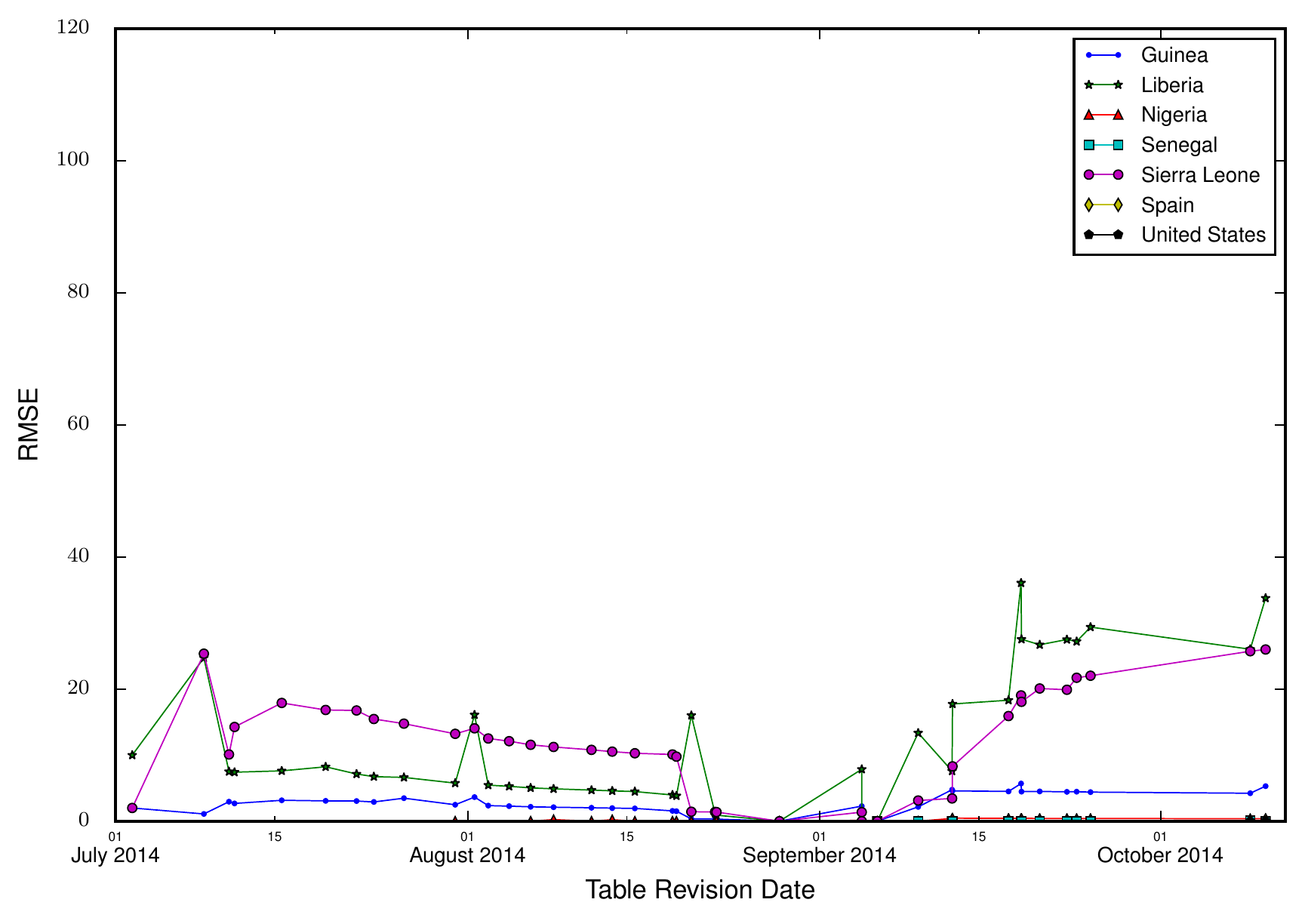}
		\caption{Deaths}
        \label{figure:deaths_rmse}
	\end{subfigure}
	\caption{Root-mean-square error (RMSE) values for the cases and deaths time series are shown for each revision where the tables changed. The RMSE spikes on July 8, 2014 (Liberia and Sierra Leone) and August 20, 2014 (Liberia) in \ref{figure:cases_rmse} were due to Wikipedia contributor errors and were fixed shortly after they were made. Most RMSE spikes are quickly followed by a decrease; this is due to updated WHO data or contributor error detection.}
	\label{figure:rmse_graphs}
\end{figure}

We compared the 39 Wikipedia epidemic time series to the ground truth data by computing the root-mean-square error (RMSE). We use the RMSE rather than the mean-square error (MSE) because the testing and ground truth time series both have the same units (cases or deaths); when they have the same units, the computed RMSE also has the same unit, which makes it easily interpretable. The RMSE,
\begin{equation}
\mathrm{RMSE} = \sqrt{\frac{1}{n} \cdot \sum_{i=1}^{n} (\hat{Y_i} - Y_i)^2},
\end{equation}
computes the average number of cases or deaths difference between a Wikipedia epidemic time series ($\hat{Y}$) and the ground truth time series ($Y$). Figure~\ref{figure:rmse_graphs} shows how the case time series and death time series RMSE changes with each table revision for each country. Of particular interest is the large spike in Figure~\ref{figure:cases_rmse} on July 8, 2014 in Liberia and Sierra Leone. Shortly after the 6:27pm spike, an edit from a different user at 8:16pm the same day with edit summary ``correct numbers in wrong country columns" corrected the error.

The average RMSE values for each country's time series are listed in Table~\ref{table:average_rmse}. Even in the worst case, the average deviation between the Wikipedia time series and the ground truth is approximately 19 cases and 12 deaths. Considering the magnitude of the number of cases (e.g., approximately 1,500 in Liberia and 3,500 in Sierra Leone during the time period considered) and deaths (e.g., approximately 850 in Liberia and 1,200 in Sierra Leone), the Wikipedia time series are generally within 1--2\% of the ground truth data.

\begin{table}
	\centering
	\small
	\caption{Average cases and deaths RMSE across all table revisions.}
	\begin{tabular}{| l | l | l | l |}
		\hline
		\textbf{Country} & \textbf{Mean Cases RMSE} & \textbf{Mean Deaths RMSE} \\ \hline
		Guinea & 3.790 & 2.701 \\ \hline
		Liberia & 18.168 & 11.983 \\ \hline
		Nigeria & 0.310 & 0.189 \\ \hline
		Senegal & 0.403 & 0.008 \\ \hline
		Sierra Leone & 18.847 & 12.015 \\ \hline
		Spain & 18.243 & 0.050 \\ \hline
		United States & 0.174 & 0.000 \\ \hline
	\end{tabular}
	\label{table:average_rmse}
\end{table}

\section{Conclusions}

Internet data are becoming increasingly important for disease surveillance because they address some of the existing challenges, such as the reporting lags inherent in traditional disease surveillance data, and they can also be used to detect and monitor emerging diseases. Additionally, internet data can simplify global disease data collection. Collecting disease data is a formidable task that often requires browsing websites written in an unfamiliar language, and data are specified in a number of formats ranging from well-formed spreadsheets to unparseable PDF files containing low resolution images of tables. Although several popular internet-based systems exist to help overcome some of these traditional disease surveillance system weaknesses, most notably HealthMap~\cite{Freifeld2008} and Google Flu Trends~\cite{Ginsberg2009}, no such system exists that relies solely on open data and runs using 100\% open source code.

Previous work explored Wikipedia access logs to tackle some of the disadvantages traditional disease surveillance systems face~\cite{McIver2014,Generous2014}. This study explores a new facet of Wikipedia: the content of disease-related articles. We present methods on how to elicit data that can potentially be used for near-real-time disease surveillance purposes. We argue that in some instances, Wikipedia may be viewed as a centralized crowdsourced data repository.

First, we demonstrate using a named-entity recognizer (NER) how case counts, death counts, and hospitalization counts can be tagged in the article narrative. Our NER, trained on a dataset derived from 14 Wikipedia articles on disease outbreaks/epidemics, achieved an F1 score of 0.753, evidence that this method is fully capable of recognizing these entities in text. Second, we analyzed the quality of tabular data available in the 2014 West Africa Ebola virus disease article. By computing the root-mean-square error (RMSE), we show that the Wikipedia time series very closely align with WHO-based ground truth data.

There are many future directions for this work. First and foremost, more training data are necessary for an operational system in order to improve precision and recall. There are many more disease- and outbreak-related Wikipedia articles that can be annotated. Additionally, other open data sources, such as ProMED-mail, might be used to enhance the model. Second, a thorough analysis of the quality and correctness of the entities tagged by the NER is needed. This study presents the methods by which disease-related named entities can be recognized, but we have not throughly studied the correctness and timeliness of the data. Third, our analysis of tabular data consisted of a single article. A more rigorous study looking at the quality of tabular data in more articles is necessary. Finally, the work presented here considers only the English Wikipedia. NERs are capable of tagging entities in a variety of other languages; more work is needed to understand the quality of data available in the 286 non-English Wikipedias.

There are several limitations to this work. First, the ground truth time series we used to compute RMSEs is static, while the Wikipedia time series vary over time. Because the relatively recent static ground truth time series may contain corrections for reporting errors made earlier in the epidemic, the RMSE values may be artificially inflated in some instances. Second, we are ignoring the user-provided edit summary. This edit summary provides information about why the edit was made. The edit summary identifies article vandalism (and subsequent vandalism reversion) as well as content corrections and updates. Taking these edit summaries into account can further improve model performance (e.g., processing edit summaries would allow us to disregard the erroneous edit that caused the July 8, 2014 spike in Figure~\ref{figure:cases_rmse}).

Ultimately, we envision this work being incorporated into a community-driven open-source emerging disease detection and monitoring system. Wikipedia access log time series gauge public interest and, in many cases, correlate very well with disease incidence. A community-driven effort to improve global disease surveillance data is imminent, and Wikipedia can play a crucial role in realizing this need.

\section{Acknowledgments}

This work is supported in part by NIH/NIGMS/MIDAS under grant U01-GM097658-01 and the DTRA Joint Science and Technology Office for Chemical and Biological Defense under project numbers CB3656 and CB10007. LANL is operated by Los Alamos National Security, LLC for the Department of Energy under contract DE-AC52-06NA25396.

\bibliography{refs,refs_extra}

\begin{thebibliography}{}

\bibitem[\protect\citeauthoryear{Adams \bgroup et al\mbox.\egroup
  }{2013}]{Adams2013}
Adams, D.~A.; Gallagher, K.~M.; Jajosky, R.~A.; Kriseman, J.; Sharp, P.;
  Anderson, W.~J.; Aranas, A.~E.; Mayes, M.; Wodajo, M.~S.; Onweh, D.~H.; and
  Abellera, J.~P.
\newblock 2013.
\newblock {Summary of Notifiable Diseases, United States, 2011}.
\newblock Technical Report~53, Centers for Disease Control and Prevention,
  Atlanta, Georgia.

\bibitem[\protect\citeauthoryear{Aramaki, Maskawa, and
  Morita}{2011}]{Aramaki2011}
Aramaki, E.; Maskawa, S.; and Morita, M.
\newblock 2011.
\newblock {Twitter catches the flu: detecting influenza epidemics using
  Twitter}.
\newblock In {\em Proceedings of the 2011 Conference on Emperical Methods in
  Natural Language Processing},  1568--1576.
\newblock Edinburgh, United Kingdom: Association for Computational Linguistics.

\bibitem[\protect\citeauthoryear{Burkhead and Maylahn}{2000}]{Burkhead2000}
Burkhead, G.~S., and Maylahn, C.~M.
\newblock 2000.
\newblock {State and Local Public Health Surveillance}.
\newblock In Teutsch, S.~M., and Churchill, R.~E., eds., {\em Principles and
  Practice of Public Health Surveillance}. New York: Oxford University Press,
  2nd edition.
\newblock chapter~12,  253--286.

\bibitem[\protect\citeauthoryear{{Centers for Disease Control and
  Prevention}}{2014}]{CDC2014_Ebola}
{Centers for Disease Control and Prevention}.
\newblock 2014.
\newblock {Ebola (Ebola Virus Disease)}.
\newblock http://www.cdc.gov/vhf/ebola/.
\newblock Accessed: 2014-10-27.

\bibitem[\protect\citeauthoryear{Culotta}{2010}]{Culotta2010}
Culotta, A.
\newblock 2010.
\newblock {Towards detecting influenza epidemics by analyzing Twitter
  messages}.
\newblock In {\em Proceedings of the First Workshop on Social Media Analytics},
   115--122.
\newblock Washington, DC: ACM Press.

\bibitem[\protect\citeauthoryear{Finkel, Grenager, and
  Manning}{2005}]{Finkel2005}
Finkel, J.~R.; Grenager, T.; and Manning, C.
\newblock 2005.
\newblock {Incorporating non-local information into information extraction
  systems by Gibbs sampling}.
\newblock In {\em Proceedings of the 43rd Annual Meeting on Association for
  Computational Linguistics}, number June,  363--370.
\newblock Morristown, NJ, USA: Association for Computational Linguistics.

\bibitem[\protect\citeauthoryear{Freifeld \bgroup et al\mbox.\egroup
  }{2008}]{Freifeld2008}
Freifeld, C.~C.; Mandl, K.~D.; Reis, B.~Y.; and Brownstein, J.~S.
\newblock 2008.
\newblock {HealthMap: Global Infectious Disease Monitoring through Automated
  Classification and Visualization of Internet Media Reports}.
\newblock {\em Journal of the American Medical Informatics Association}
  15(2):150--157.

\bibitem[\protect\citeauthoryear{Generous \bgroup et al\mbox.\egroup
  }{2014}]{Generous2014}
Generous, N.; Fairchild, G.; Deshpande, A.; {Del Valle}, S.~Y.; and
  Priedhorsky, R.
\newblock 2014.
\newblock {Global Disease Monitoring and Forecasting with Wikipedia}.
\newblock {\em PLOS Computational Biology} 10(11):e1003892.

\bibitem[\protect\citeauthoryear{Ginsberg \bgroup et al\mbox.\egroup
  }{2009}]{Ginsberg2009}
Ginsberg, J.; Mohebbi, M.~H.; Patel, R.~S.; Brammer, L.; Smolinski, M.~S.; and
  Brilliant, L.
\newblock 2009.
\newblock {Detecting influenza epidemics using search engine query data}.
\newblock {\em Nature} 457(7232):1012--1014.

\bibitem[\protect\citeauthoryear{Greenemeier}{2014}]{Greenemeier2014}
Greenemeier, L.
\newblock 2014.
\newblock {Smart Machines Join Humans in Tracking Africa Ebola Outbreak}.
\newblock {\em Scientific American}.

\bibitem[\protect\citeauthoryear{Keegan, Gergle, and
  Contractor}{2011}]{Keegan2011}
Keegan, B.; Gergle, D.; and Contractor, N.
\newblock 2011.
\newblock {Hot off the wiki: dynamics, practices, and structures in Wikipedia's
  coverage of the Tōhoku catastrophes}.
\newblock In {\em Proceedings of the 7th International Symposium on Wikis and
  Open Collaboration},  105--113.
\newblock Mountain View, California: ACM.

\bibitem[\protect\citeauthoryear{Keegan, Gergle, and
  Contractor}{2013}]{Keegan2013}
Keegan, B.; Gergle, D.; and Contractor, N.
\newblock 2013.
\newblock {Hot Off the Wiki: Structures and Dynamics of Wikipedia's Coverage of
  Breaking News Events}.
\newblock {\em American Behavioral Scientist} 57(5):595--622.

\bibitem[\protect\citeauthoryear{Keegan}{2013}]{Keegan2013a}
Keegan, B.~C.
\newblock 2013.
\newblock {A history of newswork on Wikipedia}.
\newblock In {\em Proceedings of the 9th International Symposium on Open
  Collaboration},  7:1--7:10.
\newblock Hong Kong, China: ACM.

\bibitem[\protect\citeauthoryear{Losos}{1996}]{Losos1996}
Losos, J.~Z.
\newblock 1996.
\newblock {Routine and sentinel surveillance methods}.
\newblock {\em Eastern Mediterranean Health Journal} 2(1):46--50.

\bibitem[\protect\citeauthoryear{Madoff}{2004}]{Madoff2004}
Madoff, L.~C.
\newblock 2004.
\newblock {ProMED-mail: An Early Warning System for Emerging Diseases}.
\newblock {\em Clinical Infectious Diseases} 39(2):227--232.

\bibitem[\protect\citeauthoryear{Manning, Raghavan, and
  Sch\"{u}tze}{2009}]{Manning2009}
Manning, C.~D.; Raghavan, P.; and Sch\"{u}tze, H.
\newblock 2009.
\newblock {\em {Introduction to Information Retrieval}}.
\newblock Number~c. Cambridge, England: Cambridge University Press.

\bibitem[\protect\citeauthoryear{McIver and Brownstein}{2014}]{McIver2014}
McIver, D.~J., and Brownstein, J.~S.
\newblock 2014.
\newblock {Wikipedia Usage Estimates Prevalence of Influenza-Like Illness in
  the United States in Near Real-Time}.
\newblock {\em PLOS Computational Biology} 10(4):e1003581.

\bibitem[\protect\citeauthoryear{Ng and Jordan}{2002}]{Ng2002}
Ng, A., and Jordan, M.
\newblock 2002.
\newblock {On discriminative vs. generative classifiers: A comparison of
  logistic regression and naive Bayes}.
\newblock In Dietterich, T.~G.; Becker, S.; and Ghahramani, Z., eds., {\em
  Proceedings of the 2001 Neural Information Processing Systems Conference},
  841--848.
\newblock British Columbia, Canada: MIT Press.

\bibitem[\protect\citeauthoryear{Paul and Dredze}{2011}]{Paul2011}
Paul, M.~J., and Dredze, M.
\newblock 2011.
\newblock {You are what you Tweet: Analyzing Twitter for public health}.
\newblock In {\em Proceedings of the Fifth International AAAI Conference on
  Weblogs and Social Media},  265--272.

\bibitem[\protect\citeauthoryear{Polgreen \bgroup et al\mbox.\egroup
  }{2008}]{Polgreen2008}
Polgreen, P.~M.; Chen, Y.; Pennock, D.~M.; and Nelson, F.~D.
\newblock 2008.
\newblock {Using Internet Searches for Influenza Surveillance}.
\newblock {\em Clinical Infectious Diseases} 47(11):1443--1448.

\bibitem[\protect\citeauthoryear{Signorini, Segre, and
  Polgreen}{2011}]{Signorini2011}
Signorini, A.; Segre, A.~M.; and Polgreen, P.~M.
\newblock 2011.
\newblock {The Use of Twitter to Track Levels of Disease Activity and Public
  Concern in the U.S. during the Influenza A H1N1 Pandemic}.
\newblock {\em PLOS ONE} 6(5):e19467.

\bibitem[\protect\citeauthoryear{Sutton}{2011}]{Sutton2011}
Sutton, C.
\newblock 2011.
\newblock {An Introduction to Conditional Random Fields}.
\newblock {\em Foundations and Trends® in Machine Learning} 4(4):267--373.

\bibitem[\protect\citeauthoryear{{Tjong Kim Sang} and {De
  Meulder}}{2003}]{TjongKimSang2003}
{Tjong Kim Sang}, E.~F., and {De Meulder}, F.
\newblock 2003.
\newblock {Introduction to the CoNLL-2003 shared task: language-independent
  named entity recognition}.
\newblock In {\em Proceedings of the Seventh Conference on Natural Language
  Learning at HLT-NAACL 2003 - Volume 4}, volume~4,  142--147.
\newblock Morristown, NJ, USA: Association for Computational Linguistics.

\bibitem[\protect\citeauthoryear{Toutanova \bgroup et al\mbox.\egroup
  }{2003}]{Toutanova2003}
Toutanova, K.; Klein, D.; Manning, C.~D.; and Singer, Y.
\newblock 2003.
\newblock {Feature-rich part-of-speech tagging with a cyclic dependency
  network}.
\newblock In {\em Proceedings of the 2003 Conference of the North American
  Chapter of the Association for Computational Linguistics on Human Language
  Technology - Volume 1},  173--180.
\newblock Morristown, NJ, USA: Association for Computational Linguistics.

\bibitem[\protect\citeauthoryear{{Wikimedia
  Foundation}}{2014a}]{Wikimedia2014_Wikipedia_MERS_outbreak}
{Wikimedia Foundation}.
\newblock 2014a.
\newblock {2012 Middle East respiratory syndrome coronavirus outbreak}.
\newblock
  http://en.wikipedia.org/w/index.php?title=2012\_Middle\_East\_respiratory\_syndrome\_coronavirus\_outbreak\&oldid=628796140.
\newblock Accessed: 2014-10-10.

\bibitem[\protect\citeauthoryear{{Wikimedia
  Foundation}}{2014b}]{Wikimedia2014_Wikipedia_Ebola_epidemic}
{Wikimedia Foundation}.
\newblock 2014b.
\newblock {Ebola virus epidemic in West Africa}.
\newblock
  http://en.wikipedia.org/w/index.php?title=Ebola\_virus\_epidemic\_in\_West\_Africa\&oldid=629094432.
\newblock Accessed: 2014-10-10.

\bibitem[\protect\citeauthoryear{{Wikimedia
  Foundation}}{2014c}]{Wikimedia2014_Wikipedia_Ebola_epidemic_original}
{Wikimedia Foundation}.
\newblock 2014c.
\newblock {Ebola virus epidemic in West Africa}.
\newblock
  https://en.wikipedia.org/w/index.php?title=Ebola\_virus\_epidemic\_in\_West\_Africa\&oldid=601868739.
\newblock Accessed: 2014-03-24.

\bibitem[\protect\citeauthoryear{{Wikimedia
  Foundation}}{2014d}]{Wikimedia2014_English_Wikipedia}
{Wikimedia Foundation}.
\newblock 2014d.
\newblock {English Wikipedia}.
\newblock
  http://en.wikipedia.org/w/index.php?title=English\_Wikipedia\&oldid=627512912.
\newblock Accessed: 2014-10-07.

\bibitem[\protect\citeauthoryear{{Wikimedia
  Foundation}}{2014e}]{Wikimedia2014_Wikipedia_NER}
{Wikimedia Foundation}.
\newblock 2014e.
\newblock {Named-entity recognition}.
\newblock
  http://en.wikipedia.org/w/index.php?title=Named-entity\_recognition\&oldid=627138157.
\newblock Accessed: 2014-10-11.

\bibitem[\protect\citeauthoryear{{Wikimedia
  Foundation}}{2014f}]{Wikimedia2014_Wikipedia_Wikipedia}
{Wikimedia Foundation}.
\newblock 2014f.
\newblock {Wikipedia}.
\newblock https://en.wikipedia.org/w/index.php?title=Wikipedia\&
  oldid=636552708.
\newblock Accessed: 2014-12-04.

\bibitem[\protect\citeauthoryear{{Wikimedia
  Foundation}}{2014g}]{Wikimedia2014_Wikipedia_statistics}
{Wikimedia Foundation}.
\newblock 2014g.
\newblock {Wikipedia Statistics}.
\newblock http://stats.wikimedia.org/EN/Sitemap.htm.
\newblock Accessed: 2014-10-07.

\bibitem[\protect\citeauthoryear{{Wikimedia
  Foundation}}{2014h}]{Wikimedia2014_Wikipedia_database_download}
{Wikimedia Foundation}.
\newblock 2014h.
\newblock {Wikipedia:Database download}.
\newblock http://en.wikipedia.org/w/index.php?title=Wikipedia:
  Database\_download\&oldid=627253774.
\newblock Accessed: 2014-10-08.

\bibitem[\protect\citeauthoryear{{World Health
  Organization}}{2014a}]{WHO2014_Ebola}
{World Health Organization}.
\newblock 2014a.
\newblock {Ebola virus disease}.
\newblock http://www.who.int/mediacentre/factsheets/fs103/en/.
\newblock Accessed: 2014-10-27.

\bibitem[\protect\citeauthoryear{{World Health
  Organization}}{2014b}]{WHO2014_Ebola_announcement}
{World Health Organization}.
\newblock 2014b.
\newblock {Ebola virus disease in Guinea (Situation as of 25 March 2014)}.
\newblock
  http://www.afro.who.int/en/clusters-a-programmes/dpc/epidemic-a-pandemic-alert-and-response/outbreak-news/4065-ebola-virus-disease-in-guinea-25-march-2014.html.
\newblock Accessed: 2014-12-01.

\end{thebibliography}
\bibliographystyle{aaai}

\end{document}